\documentclass[twocolumn, amssymb, nobibnotes, aps, amsmath, amssymb, superscriptaddress,floatfix,pre]{revtex4-2}
\usepackage{graphicx}
\usepackage{dcolumn}
\usepackage{bm}
\usepackage[linktocpage,colorlinks=true,linkcolor=blue,citecolor=blue,breaklinks=true,urlcolor=blue]{hyperref}
\usepackage{float}
\usepackage[utf8]{inputenc}
\usepackage[T1]{fontenc}
\usepackage{newtxtext, newtxmath}  
\DeclareGraphicsExtensions{.pdf,.jpeg,.jpg, .png, .eps, .tiff}
\usepackage{epstopdf}
\usepackage{lipsum}
\usepackage{natbib}
\usepackage[dvipsnames]{xcolor}
\usepackage[normalem]{ulem}
\usepackage{soul}
\usepackage{enumitem}
\usepackage{siunitx}
\usepackage{lineno}
\usepackage{subfigure}
\usepackage{anyfontsize}
\renewcommand{\vec}[1]{\underline{#1}}

\newcommand{\vecP}[2]{{#1}_{#2}}
\newcommand{\tens}[1]{\underline{\underline{#1}}}
\newcommand{\kbt}{k_\text{B}T}
\newcommand{\VCO}{\Xi}
\newcommand{\OCO}{\Psi_c} 

\newcommand{\av}[1]{\left\langle #1 \right\rangle}
\newcommand{\abs}[1]{\left| #1 \right|}
\newcommand{\Brac}[1]{\left( #1 \right)}

\newcommand{\order}[1]{\mathcal{O}\left( #1 \right)}

\newcommand{\fig}[1]{\textbf{Fig.~\ref{#1}}}
\newcommand{\eq}[1]{\textbf{Eq.~\ref{#1}}}

\newcommand{\sctn}[1]{\textbf{\S~\ref{#1}}}

\newcommand{\EQ}[1]{\textbf{Equation~\ref{#1}}}

\newcommand{\vel}{v}

\newcommand{\pos}{r}
\newcommand{\posV}{\vec{\pos}}
\renewcommand{\time}{t}

\newcommand{\unitT}{\tens{\delta}}

\newcommand{\act}{\zeta}

\newcommand{\rcmV}{\vec{\pos}}
\newcommand{\diff}{D}
\newcommand{\diffTrans}{\diff_0}
\newcommand{\diffEff}{\diff_\text{e}}
\newcommand{\chartime}{\tau}
\newcommand{\timePers}{\chartime_\text{p}}

\newcommand{\timeRelax}{\chartime_\text{r}}
\newcommand{\actSpeed}{\vel}

\newcommand{\pe}{\text{Pe}}
\newcommand{\wi}{\text{Wi}}
\newcommand{\er}{\text{Er}}
\newcommand{\noise}{\xi}
\newcommand{\noiseV}{\vec{\noise}}
\newcommand{\ori}{\vec{\hat{e}}}

\newcommand{\peclet}{P\'{e}clet }
\DeclareSIUnit{\molar}{M}

\newcommand{\ue}{School of Physics and Astronomy, The University of Edinburgh, Peter Guthrie Tait Road, Edinburgh, EH9 3FD, United Kingdom}

\definecolor{pumpkin}{rgb}{1.0,0.4,0.0}
\definecolor{midnight}{rgb}{0.003921569,0.098039216,0.576470588}
\definecolor{saphire}{rgb}{0.0,0.196,0.372549}
\definecolor{crimson}{rgb}{0.75686,0,0.262745}
\definecolor{capri}{rgb}{0.0,0.768627,0.8745098}
\definecolor{amber}{rgb}{0.95686,0.66666667,0.0}
\definecolor{plum}{rgb}{0.50588,0.007843,0.3843137}
\definecolor{cerulean}{rgb}{0.0,0.568627,0.70980}
\definecolor{ruby}{rgb}{0.83137,0.0,0.4470588}

\begin{document}
\title{Scaling laws for passive polymer dynamics in active turbulence}
\author{Zahra K. Valei}
\affiliation{\ue}
\author{Tyler N. Shendruk}
\affiliation{\ue}
\email{t.shendruk@ed.ac.uk}

\date{\today}

\begin{abstract}
    \noindent
    Biological systems commonly combine intrinsically out-of-equilibrium active components with passive polymeric inclusions to produce unique material properties.
    To explore these composite systems, idealized models---such as polymers in active fluids---are essential to develop a predictive theoretical framework.
    We simulate a single, freely jointed passive chain in two-dimensional active turbulence.
    Active flows advect the polymer, producing a substantially enhanced diffusivity.
    Our results reveal that the dimensionless diffusivity obeys scaling laws governed by the Péclet, Weissenberg, and Ericksen numbers,
    which paves the way for designing active/polymeric hybrid materials with predictable properties that differ significantly from those of nondeformable inclusions.
\end{abstract}

\maketitle


\begin{figure}[tb]
     \centering
     \includegraphics[width=\linewidth]{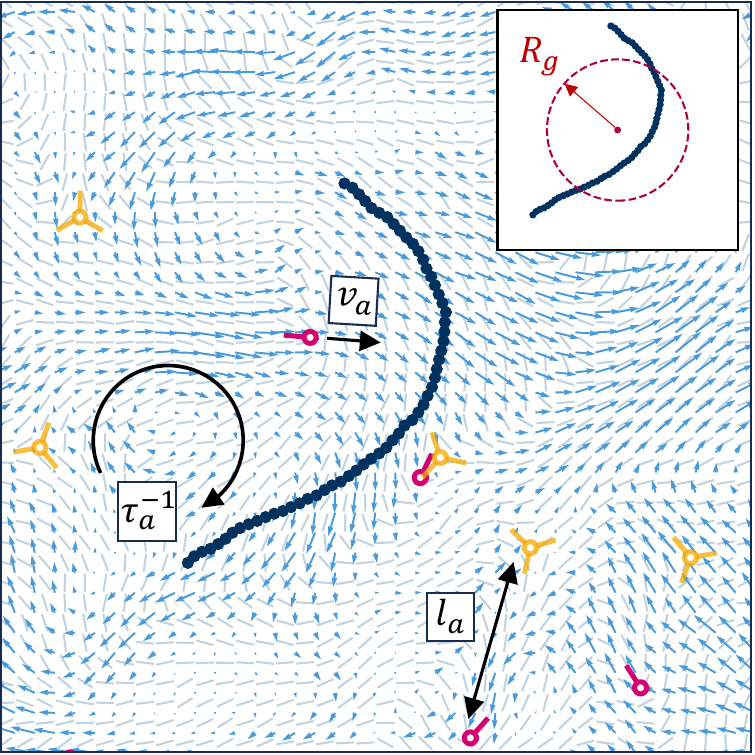}
     \caption{
     \textbf{Snapshot of a polymer in active nematic turbulence.} 
     The polymer consists of $N=60$ navy monomers. 
     The director is shown as silver lines and the flow field as blue vectors.
     Defects are depicted in ruby ($+1/2$) and amber ($-1/2$). 
     The average distance between defects is the active length scale $l_a$. 
     The active time scale $\tau_a$ is represented as the inverse of vorticity.
     The motile $+1/2$ defect speed is characterized as $v_a = l_a/\tau_a$. 
     Activity is $\zeta = 0.08$ and system size is $L=80$. 
     \textbf{Inset} Polymer center of mass and gyration radius $R_g$. 
     }
     \label{fig:schematic}
\end{figure}

The presence of both active and passive components is a hallmark of biomaterials~\cite{winkler2020physics}. 
Examples include intracellular molecular motors interacting with passive cytoskeletal filaments~\cite{kapral2016stirring,mikhailov2015hydrodynamic,lu2016microtubule}, polymerases transcribing DNA~\cite{mejia2015trigger}, and the dynamic interplay between euchromatin and heterochromatin~\cite{cremer20154d,ganai2014chromosome,mahajan2022euchromatin,eshghi2023,chan2024,saintillan2018extensile}.
In such composite biomaterials, active components are intrinsically out-of-equilibrium due to continuous energy conversion into coherent movement~\cite{bechinger2016active,thampi2015intrinsic}, while passive components are driven from equilibrium through their sustained contact with the active phase.
Previous work has focused primarily on rigid passive bodies embedded in active flows, including tracer particles in dilute suspensions of swimming bacteria~\cite{morozov2014enhanced,leptos2009dynamics,underhill2008diffusion}, colloids in dense bacterial communities~\cite{wu2000particle,harder2014activity,mino2011enhanced,gregoire2001active,valeriani2011colloids} and disks in active nematic films~\cite{ray2023rectified,neville2024controlling}.

\begin{figure*}[tb]
    \centering
    \includegraphics[width=\linewidth]{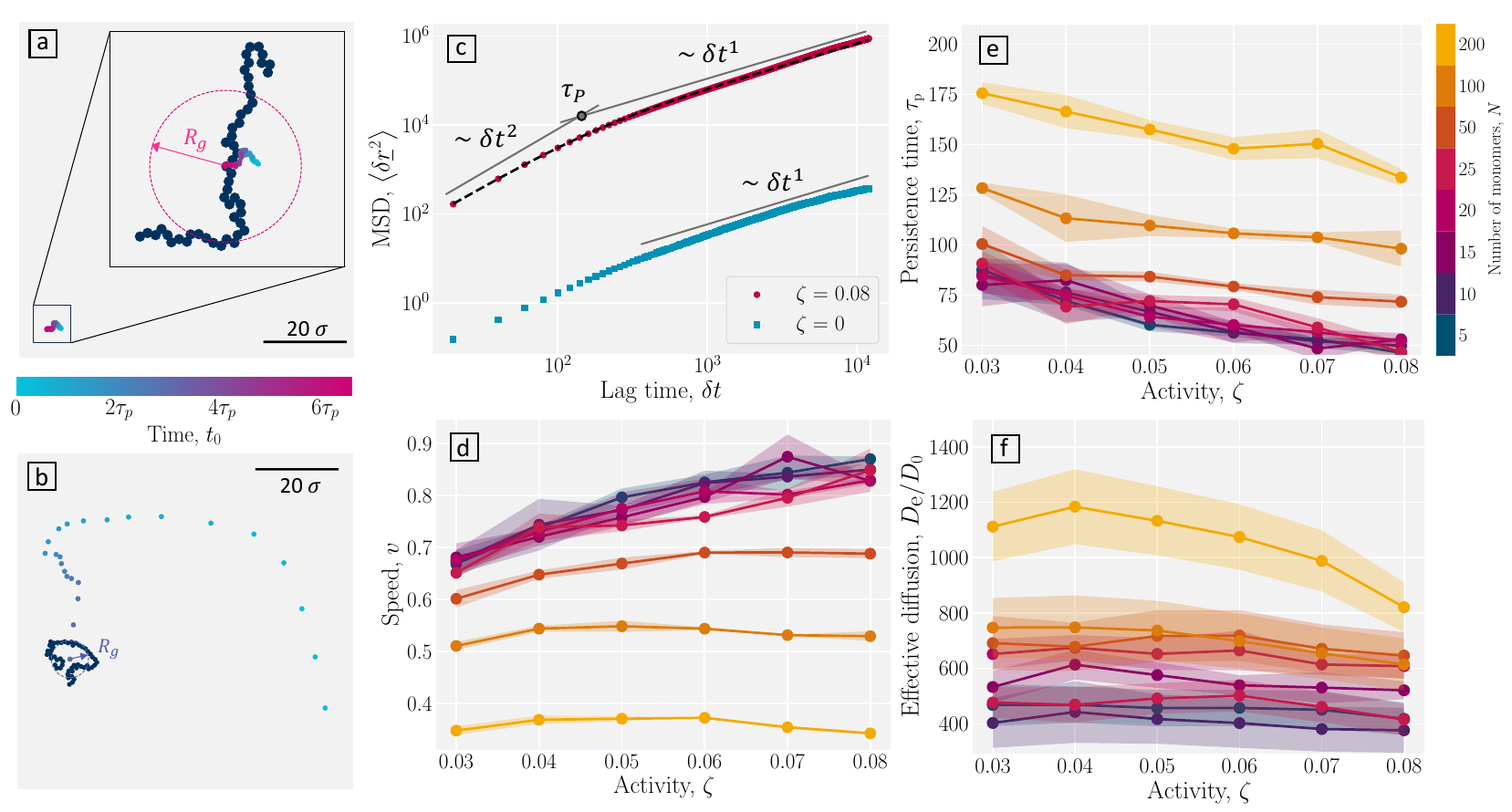}
    \caption{
    \textbf{Mean-squared displacement (MSD) of the polymers' center of mass.}
    (a), (b) Representative center-of-mass trajectories of a polymer consisting of $N = 50$ monomers of size $\sigma$, color-coded by time $t_0$.
    \textbf{(a)} Trajectory without activity ($\zeta = 0$) and \textbf{(b)} with activity ($\zeta = 0.08$).
    \textbf{Inset} A zoomed-in view of the trajectory and polymer conformation and the instantaneous gyration radius $R_g$.
    \textbf{(c)} The MSD $\av{\delta \rcmV^2}$ of the examples from (a)-(b) as a function of lag time $\delta t$, averaged over 40 realizations and longer times. 
    The passive case ($\act=0$) exhibits long-time diffusive behavior $\av{\delta \rcmV^2} \sim \delta t$.
    The active case ($\act=0.08$) shows a short-time propulsive regime $\av{\delta \rcmV^2} \sim \delta t^2$, followed by effective diffusive behavior at times longer than the persistence time $\timePers$.
    \textbf{(d)} The center-of-mass speed $v$ of polymers with $N$ monomers.
    \textbf{(e)} Persistence time of advection by the flow.
    \textbf{(f)} Effective diffusion coefficients $\diffEff$ normalized by thermal diffusivity $\diffTrans$.
    }
    \label{fig:MSD}
\end{figure*}

While previous studies have centered on rigid inclusions in active surroundings~\cite{loewe2022passive,di2010bacterial,Thampi2016,nishiguchi2018,reinken2020,velez2024,schimming2024active}, polymers and flexible filaments embedded in active backgrounds are of specific biological importance.
They are crucial for the formation of cytoplasmic flows~\cite{kimura2017endoplasmic,lu2023go,illukkumbura2020patterning,stein2021swirling,gubieda2020going}, the organization of the extracellular matrix~\cite{whitchurch2002extracellular}, and the transport of DNA~\cite{mccuskey2024}.
Moreover, macromolecules are commonly found in bacterial environments, including gastrointestinal tracts, mucus, blood, and saliva~\cite{liao2023viscoelasticity,whitchurch2002extracellular}.
Such systems are especially intriguing due to the interplay between activity and conformational entropy~\cite{weady2024}. 
Prior work has mainly explored the conformational changes of semi-flexible polymers in ensembles of self-propelled particles~\cite{kaiser2014unusual,shin2015facilitation,mousavi2021active} or athermal baths~\cite{eisenstecken2017conformational,eisenstecken2016conformational,eisenstecken2017internal,samanta2016chain,ghosh2014dynamics,vandebroek2015dynamics,sakaue2017active}.
These systems illustrate how intrinsically out-of-equilibrium surroundings can actively deform passive polymeric inclusions with internal degrees of freedom, yet they do not account for the hydrodynamics of active fluids, nor the ensuing active turbulence.
These spontaneous flows play a vital role in intracellular mixing and transport~\cite{chakrabarti2024cytoplasmic,htet2023cortex,htet2025analytical,drechsler2020optical}, as well as macromolecular nutrient distribution, chemical signal dispersal and molecular waste removal in active microbial communities~\cite{cisneros2011dynamics,wensink2012meso,wheeler2019not,kurtuldu2011enhancement}.
While the enhanced transport in active flows is widely recognized ~\cite{granek2022anomalous,angelani2010geometrically,kaiser2012capture,kaiser2013capturing,kaiser2014transport,mallory2014curvature,angelani2009self}, a predictive theoretical framework for the active transport of polymeric inclusions is lacking.

We study the dynamics of a single 
passive, freely jointed polymer embedded in active nematic (\fig{fig:schematic}). 
Active nematics in bulk exhibit active turbulence, a low Reynolds number phenomenon with characteristic length and time scales (\fig{fig:schematic})~\cite{shankar2022topological,alert2022active}.
This composite system exemplifies the competition between entropy-driven relaxation processes and activity-driven dynamics. 
We employ a hybrid simulation approach, combining active nematics multi-particle collision dynamics to model the active nematic background (\sctn{sctn:mpcd}) and molecular dynamics to simulate the polymer, which is coupled directly to the velocity but not the orientation field (\sctn{sctn:md}). 
This approach captures fluctuations through the thermal energy $\kbt$, hydrodynamic interactions, nematic elasticity through $K$ and active stresses via the activity $\act$.
We find that passive polymers within active nematic turbulence experience a thousand-fold increase in diffusivity, arising from short-time advection of the polymers by active flows with a finite correlation time.
This effect is characterized by a pair of dimensionless numbers: 
The \peclet number $\pe$, quantifying polymer advection rate relative to the diffusion rate, and the Weissenberg number $\wi$, comparing the deformation rate with relaxation time. 
These are shown to be linked via the Eriksen number $\er$, representing the ratio of characteristic length scales.
The interplay between these dimensionless numbers fully determines the enhanced diffusivity of passive polymers in active turbulence.

This study focuses on flow-aligning nematics with extensile activity ($\zeta>0$) and values are reported in simulation units (\sctn{sctn:unitsParameters}). 
The competition between nematic elasticity $K$, viscosity $\eta$ and activity $\act$ produce active length and time scales (\sctn{appendix:turb})~\cite{guillamat2017taming,hemingway2016correlation,gulati2022boundaries}
\begin{align}
    l_a \sim \sqrt{\frac{K}{\act}} 
    \qquad &; \qquad
    \tau_a \sim \frac{\eta}{\act}
    \label{eq:actLengthTime}
\end{align} 
over which flow remains coherent in active turbulence. 

The polymer dynamics are characterized by the mean squared displacement (MSD) of its center of mass $\av{\delta \rcmV^2} = \av{ \Brac{ \rcmV\Brac{t+\delta t} - \rcmV\Brac{t} }^2 }$, where $\rcmV$ is the center of mass position and $\av{\cdot}$ is the ensemble and temporal average over all starting times $t$ (\sctn{sctn:msd}). 
Without activity, the polymer interacts with the thermal fluctuations and randomly explores space (\fig{fig:MSD}a), leading to diffusive motion $\av{\delta \rcmV^2} = 2d \diffTrans \delta t$, where $\diffTrans$ is the thermal diffusion coefficient in $d$ dimensions (\fig{fig:MSD}c). 
This defines $\timeRelax=R_g^2/\diffTrans$, the polymer relaxation time over which
it diffuses a distance equal to its gyration radius $R_g$ (\fig{fig:schematic}, inset).

In active turbulence, the polymer explores a larger region than its passive counterpart (\fig{fig:MSD}b).
This is due to short-time advection by the active flows, where $\av{\delta \rcmV^2} = v^2 \delta t^2$ with the center of mass propulsion speed of polymer $v$ (\fig{fig:MSD}c).  
The propulsion speed increases with activity for short polymers but saturates at high activities (\fig{fig:MSD}d).
For long polymers ($N \geq 50$), it decreases and becomes relatively independent of activity. 
However, the turbulent nature of these active flows decorrelates the advective motion. 
The time over which the polymer is persistently advected by active flows is $\timePers$ (\fig{fig:MSD}c; \sctn{sctn:msd}).
For all polymer lengths, $\timePers$ decreases monotonically with activity, while increasing polymer length leads to longer propulsion times (\fig{fig:MSD}e).

Together, the advective polymer speed $v$ and the persistence time $\timePers$ produce the long-time enhanced diffusion coefficient (\fig{fig:MSD}c; \sctn{sctn:msd})
\begin{align}
    \diffEff &= \diffTrans + \frac{v^2 \timePers}{d} .
    \label{eq:effDiff}
\end{align}
This persistent dynamics model accurately describes the MSD of polymers in active turbulence (\fig{fig:MSD}c; dashed line).
All polymers show a significant increase in their effective diffusivity compared $\diffTrans$ (\fig{fig:MSD}f).
The enhanced diffusivity relative to the thermal diffusivity increases with length $N$. 
While polymers with $N < 200$ monomers maintain a relatively constant diffusivity, the $N=200$ polymer exhibits nonmonotic diffusivity as activity rises.
The observed diffusivity is $\order{10^3}$ times greater than the thermal diffusivity for $N=200$. 
This enhancement arises from the active kicks the polymers receive.
As activity increases, the fluid speed rises; however, the coherent time of the flows decreases, corresponding to a reduction in the active time scale $\tau_a$.
These counteracting effects of increased speed and reduced coherent time result in relatively constant effective diffusion coefficients (\fig{fig:MSD}f).

We have seen that the speed $v$, persistence time $\timePers$ and, consequently, effective diffusivity $\diffEff$ vary with the activity of the surrounding fluid $\act$ and the polymer length $N$. 
For macromolecules embedded in a flowing fluid, the dimensionless \peclet number ($\pe$) quantifies the importance of advection relative to thermal diffusion ~\cite{bianco2018globulelike,philipps2022dynamics} as
\begin{align}
    \pe &= \frac{v/R_g}{D_0/R_g^2} = \frac{vR_g}{\diffTrans},
    \label{eq:pe}
\end{align}
where $R_g$ is the gyration radius of the polymer in equilibrium, which in $d=2$ dimensions~\cite{shannon1997dynamical,koelman1990cellular} is verified numerically (\sctn{sctn:passive}) to scale as
\begin{align}
    R_g &\sim N^{3/4}.
    \label{eq:rg}
\end{align}
Computational studies in 2D have demonstrated the theoretical expectation for the diffusion coefficient is
\begin{align}
    D_0 = \frac{\kbt}{4\pi\eta} \left[ \ln \frac{l_\text{sys}}{R_g} + \text{const} \right] \equiv \mathcal{D}^{-1}_N , 
    \label{eq:diff2D}
\end{align}
where $l_\text{sys}$ is the system size (\sctn{sctn:passive})~\cite{falck2003dynamics,punkkinen2005dynamics}.
The change of variables to inverse diffusion $\mathcal{D}_N$ highlights the logarithmic dependence on $N$ via $R_g$ (\sctn{sctn:preliminaryPe}). 

\begin{figure*}[tb]
    \centering
    \includegraphics[width=\linewidth]{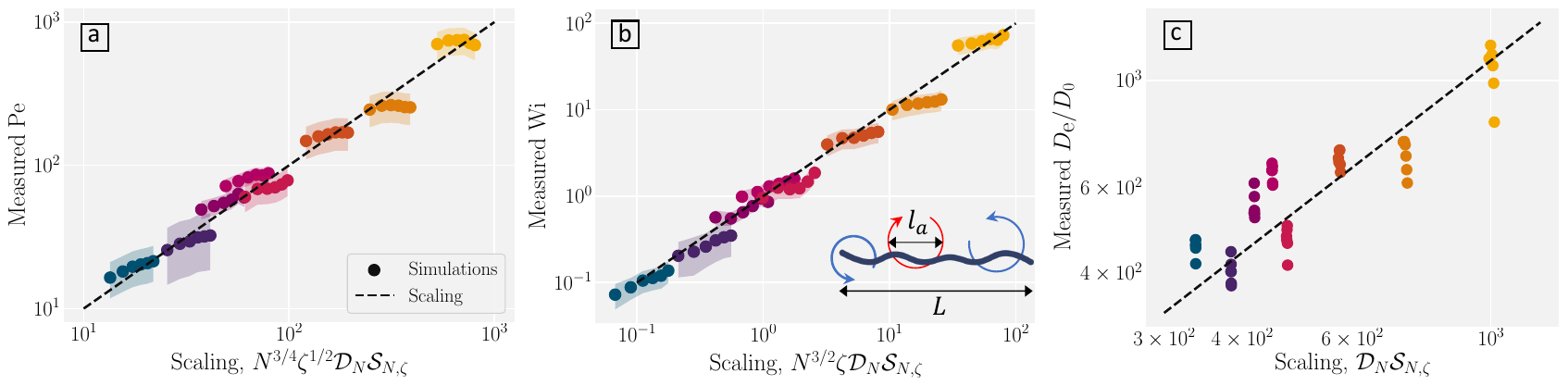}
    \caption{
    \textbf{Dynamics of polymers expressed in terms of dimensionless numbers and relevant scaling.} 
    \textbf{(a)} \peclet number $\pe$. 
    \textbf{(b)} Weissenberg number $\wi$. 
    \textbf{(c)} Effective diffusion $\diffEff$ normalized by thermal diffusivity $\diffTrans$. 
    The y-axes are measured from simulations and the x-axes are the scaling predictions for polymers of length $N$ in active turbulence with activity $\act$. 
    $\mathcal{D}_N$ is the inverse of the thermal diffusion coefficient and $\mathcal{S}_{N,\zeta}$ is the crossover factor on the polymer velocity.
    \textbf{Inset} The polymer contour length $L$ can be subdivided into $n = L/l_a$ segments which are as long as the activity length scale $l_a$. 
    }
    \label{fig:Dimensionless}
\end{figure*}

The polymer's advection speed $v$ is hypothesized to be proportional to the characteristic active flow speed $v_a$, which itself goes as 
\begin{align}
    v_a &\sim \frac{l_a}{\tau_a} \sim \act^{1/2}
\end{align}
by~\eq{eq:actLengthTime}~\cite{hemingway2016correlation}. 
Moreover, the polymer's speed $v$ exhibits a length-dependent behavior (\fig{fig:MSD}d).
Specifically, shorter polymers ($N\leq50$) show increasing speed with activity, while longer polymers exhibit a saturation, with their speed generally decreasing with length.
To explain the lower speeds of longer polymers, consider a polymer that can span many decorrelated regions of active turbulence (\fig{fig:Dimensionless} inset).
Such a long polymer interacts with multiple ``patches'', each of which is a coherent flow, but decorrelated from neighboring patches. 
The cumulative effect of these randomly misaligned flows reduces the average velocity of the polymer's center of mass, explaining the decline in enhanced diffusivity for long polymers ($N=200$) and high activities (\fig{fig:MSD}f).
To model this effect, consider a sufficiently long polymer as a chain of $n = L/l_a \sim N\act^{1/2}$ segments, each subject to uncorrelated flow. 
For large $n$, the variance scales as $\av{v^2} \sim n$.
The center of mass speed is then the average of these across the $n$ segments, leading to $v \sim 1/\sqrt{n}$.
Hence, the polymer speed requires a correction factor $v=\mathcal{S}_{N,\zeta} v_a$. 
The function $\mathcal{S}_{N,\zeta}$ should be unity for small lengths $n \ll 1$ and cross over to scale as $n^{-1/2}$ for $n\gg1$. 
We pragmatically choose
\begin{align}
     \mathcal{S}_{N,\zeta}(\alpha,\beta;n) &= \alpha \left( 1 - \frac{n^{1/4}-\beta}{\sqrt{1+\Brac{n^{1/4}-\beta}^2}} \right) ,
     \label{eq:crossover}
\end{align} 
which has the correct limiting behaviors, where the fitting parameters $\alpha$ absorbs any nonscaling factors and $\beta$ controls the crossover point, which is expected to be of order $\order{1}$. 
The subscripts emphasize that the crossover function depends on both $N$ and $\act$ through $n$. 
In the short polymer limit, $\lim_{n\ll1}\mathcal{S}_{N,\zeta}\to\text{constant}$, which means that $v= \lim_{n\ll1}\mathcal{S}_{N,\zeta}v_a \sim \act^{1/2}$. 
On the other hand, in the long polymer limit, $\lim_{n\gg1}\mathcal{S}_{N,\zeta} \to \alpha n^{-1/2} /2$ and $v=\lim_{n\gg1}\mathcal{S}_{N,\zeta}v_a \sim \act^{1/4}N^{-1/2}$. 

This correction factor is applied to the polymer speed and, along with the scaling expectations for $R_g$ (\eq{eq:rg}) and $D_0$ (\eq{eq:diff2D}), is substituted into~\eq{eq:pe} to find
\begin{align}
    \pe &\sim N^{3/4} \zeta^{1/2} \mathcal{D}_N \mathcal{S}_{N,\zeta} .
\end{align}
The predicted scaling agrees well with measured simulation values (\fig{fig:Dimensionless}a).
The motion of all polymers is primarily governed by advection, as indicated by $\pe \gg 1$ (\fig{fig:Dimensionless}a), consistent with the large effective diffusivity (\fig{fig:MSD}).
The fitting parameter $\alpha = 0.59 \pm 0.07 $ captures the combined effects of absorbed constants, while $\beta = 2.37 \pm 0.34 \sim \order{1}$ as expected.

Having collapsed the polymer's advective velocity through the \peclet number (\fig{fig:Dimensionless}a), the same can be done for the persistent time via the Weissenberg number~\cite{poole2012deborah,dealy2010weissenberg} 
\begin{equation}
    \wi = \frac{\timeRelax}{\timePers},
    \label{eq:Weissenberg}
\end{equation}
where $\timeRelax = R^2_g / D_0$ is the thermal relaxation time of the polymer and $\timePers$ is the time scale of polymer advection.
The polymer's persistent advection time is estimated as
\begin{align}
    \tau_p &= \frac{l_a}{v} = \frac{l_a}{v_a \mathcal{S}_{N,\zeta}},
\end{align}
which is the time it takes for the polymer to explore a distance $l_a$ at speed $v$. 
The ratio of $\timeRelax$ and $\tau_p$ causes the Weissenberg number to scale as 
\begin{align}
    \wi &= \frac{R_g^2}{\diffTrans} \frac{v}{l_a} \sim N^{3/2}\zeta \mathcal{D}_N \mathcal{S}_{N,\zeta} .
\end{align}
The expected value of $\wi$ matches well with the simulations (\fig{fig:Dimensionless}b).
The shorter polymers ($N<20$) follow the expectation and relax to a relatively undeformed conformation, as indicated by $\wi<1$. 
Longer polymers with $\wi\gg1$ are substantially elongated, extending over many active length scales. 

Since the \peclet number describes advection speed and the Weissenberg number characterizes propulsion time, the effective diffusion coefficient (\eq{eq:effDiff}) can be expressed in terms of these dimensionless numbers as
\begin{align}
    \frac{\diffEff}{\diffTrans} -1 &= \frac{\pe^2}{d \wi} \sim \mathcal{D}_N \mathcal{S}_{N,\zeta}.
    \label{eq:effDiffPe}
\end{align}
Surprisingly, the power law scaling of $\pe^2$ with $N$ and $\act$ cancels with the scaling of $\wi$ in $d=2$ dimensions, leaving only the logarithmic scaling $\mathcal{D}_N$ (\eq{eq:diff2D}) and the crossover scaling $\mathcal{S}_{N,\zeta}$ (\eq{eq:crossover}). 
\EQ{eq:effDiffPe} is found to predict the effective diffusivity (\fig{fig:Dimensionless}c). 
While we have focused on $d=2$ since active nematics are typically two dimensional films~\cite{rivas2020driven,Houston2024,Bantysh2024}, the scaling in $d=3$ is expected to be $\pe\sim N^{2\nu} \act^{1/2} \mathcal{S}_{N,\zeta}$ and $\wi\sim N^{3\nu} \act \mathcal{S}_{N,\zeta}$, such that $\diffEff/\diffTrans -1 \sim N^\nu \mathcal{S}_{N,\zeta}$ for $\nu\approx3/5$.

So far, the dynamics has been described in terms of time scales (advective rate over diffusive in $\pe$ and relaxation time over persistence time in $\wi$), but this can be rephrased in terms of length scales via the Eriksen number $\er = \eta v_a R_g/K$ ~\cite{rorai2021active,kos2019mesoscopic}. 
Here, $\er$ indicates whether the active flow $v_a$ overcomes the elasticity $K$ of the nematic, disrupting its orientation on the scale of $R_g$. 
The Eriksen number is found to relate the \peclet and Weissenberg numbers as 
\begin{align}
    \er = \frac{\wi}{\pe} = \frac{R_g}{l_a} ,
    \label{eq:er}
\end{align} 
which shows that the dimensionless activity numbers commonly used in the literature can be interpreted as Ericksen numbers~\cite{chandragiri2019active,thampi2022channel,samui2021flow,keogh2022helical,head2024majorana,alert2020universal,loewe2022passive,doostmohammadi2017onset,shendruk2017dancing,shendruk2018twist}.
Therefore, the effective diffusion can be rewritten as 
\begin{align}
    \frac{\diffEff}{\diffTrans}-1 &= \frac{\wi}{d\er^2}. 
\end{align}
In fact, effective diffusivity can be directly written as a ratio of time and length scales as
\begin{align}
    \frac{\diffEff}{\diffTrans}-1 &= \frac{1}{d} \Brac{\frac{\timeRelax}{\timePers}}\Brac{\frac{l_a}{R_g}}^2. 
\end{align}

This study investigated the dynamics of single flexible polymers embedded in active nematic turbulence, finding that activity significantly enhances polymer diffusion.
The enhancement arises from the short-time advection of the polymer and decorrelation of active turbulence. 
The advection rate for shorter polymers increases with activity and is directly proportional to the characteristic velocity of the active turbulence; whereas, longer polymers reach a plateau.
From the ratio of \peclet number, which reveals the scaling for advection speed, and Weissenberg number, which characterizes the shear, the scaling for effective diffusion of polymers can be predicted. 

This study sets the groundwork for future research on polymer dynamics in active media and may have direct consequences for macromolecular transport in bacterial swarms~\cite{hallinen2025bacterial,wheeler2019not,wensink2012meso}. 
Studies of the dynamics of polymers suspended in classical inertial turbulence have revealed a wide array of dynamics~\cite{vincenzi2021effect,vincenzi2021polymer,serafini2024polymers,lin2024maximum} and the present work suggests a similar variety of dynamics may occur in active turbulence. 
The success of scaling arguments suggests that integrating polymer physics with active matter could provide valuable insights into biomaterials.
Recent studies of long polymers confined in spherical cavities with active nematics have helped to illuminate chromatin dynamics~\cite{weady2024}, dense suspensions of polymers in active fluids can serve as a model for cytoskeletal assemblies~\cite{yan2022toward,sanchez2012spontaneous,belmonte2017theory,roostalu2018determinants}, polymer brushes in contact with active surroundings may model for biosensors such as protein channels~\cite{martin2007learning,yameen2009synthetic,mei2022bioinspired,wang2021bio} and polymer gels in an active solvent may represent the cell cortex~\cite{prost2015active,turlier2014furrow,da2022viscous,svitkina2020actin}. 
Thus, studying the interplay between the polymeric degrees of freedom and active media can reveal the pathways through which biological systems orchestrate their functions.

\paragraph*{Acknowledgments}
This research has received funding from the European Research Council (ERC) under the European Union’s Horizon 2020 research and innovation program (Grant agreement No. 851196). 
We acknowledge useful discussions with Davide Marenduzzo, Gavin Melaugh, Benjam\'{\i}n Loewe, Oleksandr Baziei, Manasa Kandula.
For the purpose of open access, the author has applied a Creative Commons Attribution (CC BY) license to any Author Accepted Manuscript version arising from this submission.

\bibliographystyle{unsrt}
\bibliography{references}

\newpage
\appendix 

\section{Numerical Methods}

\subsection{Active Nematic Multi-Particle Collision Dynamics}
\label{sctn:mpcd}

To model active nematic fluids, we use active-nematic multi-particle collision dynamics (AN-MPCD)~\cite{kozhukhov2022mesoscopic}.
It is chosen for its capability to simulate embedded particles within a fluctuating active nematohydrodynamic background.

This mesoscopic simulation technique represents the fluid as point particles characterized by mass $m_i$, position $\vecP{\vec r}{i}(t)$, velocity $\vecP{\vec v}{i}(t)$, and nematic orientation $\vecP{\vec u}{i}(t)$~\cite{shendruk2015multi}.
The number of lines beneath the notation represents the tensor rank.
While time is discretized into steps of size $\Delta t$, other quantities evolve continuously.
AN-MPCD consists of two steps: the streaming step and the collision step.
In the steaming step, particle positions evolve ballistically as
\begin{align}
    \vecP{\vec r}{i}\Brac{t+\Delta t} &= \vecP{\vec r}{i}\Brac{t} + \vecP{\vec v}{i}\Brac{t} \Delta t.   
\end{align}

The collision step involves two phases: $(1)$ momentum exchange and $(2)$ orientation fluctuations.
\paragraph*{(1) Momentum exchange.} The velocity of particle $i$ within cell $c$ is updated as $\vec{v}_i(t+\Delta t) = \vec{v}^{\text{cm}}_c(t)+ \vec{\VCO}_{i,c}$, where $\vec{v}^{\text{cm}}_c(t) = \av{\vec{v}_i(t)}_c$ is the velocity of the cell center of mass at time $t$, $\av{\cdot}_c$ represents the mass-weighted average over the $N_c$ particles instantaneously within the cell $c$ and $\vec{\VCO}_{i,c}$ is the collision operator.
The collision operator consists of a passive contribution ${\vec{\VCO}}^{\text{P}}_{i,c}$ and an active part ${\vec{\VCO}}^{\text{A}}_{i,c}$.
We assume all the fluid particles have the same mass $m$. 
For the passive contribution, a modified angular-momentum conserving Andersen-thermostatted collision operator is employed. 
In the absence of angular momentum conservation, the Andersen-thermostatted collision operator is ${\vec{\VCO}}^{\text{P}}_{i,c} = {\vec{v}^{\text{ran}}_i} - \av{{\vec{v}^{\text{ran}}_i}}_c$, where the components of ${\vec{v}^{\text{ran}}_i}$ are Gaussian random numbers 
with variance $\kbt/m$ and zero mean~\cite{Noguchi2007,allahyarov2002mesoscopic}. 
Although the individual velocity of each particle is randomized, subtracting $\av{{\vec{v}^{\text{ran}}_i}}_c$ assures that the cell center-of-mass velocity---and consequently the total momentum---remains unchanged~\cite{Noguchi2007}. 
To conserve angular momentum, an additional term must be added to the collision operator to remove any angular velocity generated by the collision operation~\cite{gotze2007relevance}. 
In an isotropic system, the change in the angular momentum can rise due to the stochastically generated velocities 
$ \delta \vec{\mathcal{L}}_{\text{vel}} = m \sum_{j=1}^{N_c(t)} \vec{r}_{j,c} \times (\vec{v}_j - {\vec{v}^{\text{ran}}_j})$, where $\vec{r}_{i,c} = \vec{r}_i - \av{\vec{r}_i}_c$ is the relative position of particle $i$ with respect to the center of mass of all particles in the cell. 
In a nematic system, changes in the orientation will contribute additional angular velocity $\vec{\mathcal{L}}_{\text{ori}}$, which will be elaborated below. 
For angular-momentum conserving Andersen-thermostatted AN-MPCD the passive collision operator is
\begin{align}
    \label{eq:CollisionOperator}
    {\vec{\VCO}}^{\text{P}}_{i,c} &= {\vec{v}^{\text{ran}}_i} - \av{{\vec{v}^{\text{ran}}_i}}_c + \Brac{{\tens{I}^{-1}_c} \cdot \Brac{\delta \vec{\mathcal{L}}_{\text{vel}}+\delta\vec{\mathcal{L}}_{\text{ori}}}} \times \vec{r}_{i,c},
\end{align} 
where the moment of inertia tensor $\tens{I}_c$ is for the particles in cell $c$. 

Activity is introduced as a force dipole that locally injects energy at the scale of one MPCD cell without affecting momentum, applied by the active component of the collision operator
\begin{align}
    \label{eq:ActiveNematic}
    {\vec{\VCO}}^{\text{A}}_{i,c} &= 
    \zeta_c \Delta t \Brac{
    \frac{\kappa_i}{m_i}-\frac{\av{\kappa_i}_c}{\av{m_i}_c}
    }
    \vecP{\vec{n}}{c},
\end{align}
where $\zeta_c$ sets the strength of cellular force dipole ~\cite{kozhukhov2022mesoscopic}.
To determine the direction of the active force acting on particle $i$, cell $C$ is divided by a plane passing through its center of mass, with a normal vector $\vecP{\vec{n}}{c}$. 
Depending on whether $\vecP{\vec{r}}{i}$ lies above or below this plane, the value of $\kappa_i(\vecP{\vec{r}}{i}, \vecP{\vec{n}}{c})$ is assigned as $\pm 1$, indicating the direction of the active force.
The first term in \eq{eq:ActiveNematic} represents individual impulses per unit mass on particle $i$, and the second term removes any residual impulse to locally conserve momentum. 

The strength of the active dipole in cell $c$ can be assigned in various ways, depending on the physics of interest.
For instance, if all $N_c$ fluid particles within a cell are treated as active units, $\zeta_c = \sum_{j=1}^{N_c} \zeta_j$.
In this scheme there is a positive feedback between activity and density that can exacerbate density inhomogeneities\cite{kozhukhov2024mitigating}.
Since, density inhomogeneities affects the dynamics of inclusions~\cite{zantop2021multi}, we have chosen a modulated strength to mitigate activity-induced density fluctuations~\cite{kozhukhov2024mitigating}.
This modulated activity is given by
\begin{equation}
    \zeta_c = S_c (N_c; \sigma_p,\sigma_w) \sum_{j=1}^{N_c} \zeta_j,
\end{equation}
where $S_c$ is a sigmoidal function expressed as
\begin{equation}
    S_c (N_c; \sigma_p,\sigma_w) = \frac{1}{2}
    \Brac{
    1 - \tanh
    \Brac{
    \frac{N_c - \av{N_c} \Brac{1+\sigma_p}}{\av{N_c}\sigma_w}
    }}.
\end{equation}
This function compares a cell instantaneous number density $N_c(t)$ to the system-wide average $\av{N_c}$ and returns a value between 0 and 1.
The parameters $\sigma_p$ and $\sigma_w$ control the position and width of the sigmoid drop.
Specifically, $\sigma_p$ determines the position of the sigmoid midpoint, with $\sigma_p = 0$ placing the midpoint at $\av{N_c}$ , while $\sigma_p>0$ $(\sigma_p<0)$ shifts the midpoint to higher (lower) densities.

\paragraph*{(2) Orientation fluctuations.} 
Orientation of particle $i$ updates as $\vecP{\vec u}{i}\Brac{t+\Delta t} = \OCO$, where $\OCO$ is the nematic collision operator acting on particles in cell $c$.
The collision operator acts as a rotation, modifying the orientation of each AN-MPCD particle during the time step $\Delta t$.
The reorientation process involves (i) a \textit{stochastic} contribution $\Brac{\delta \vecP{\vec u}{i}^{ST}/\delta t}$ and (ii) a \textit{flow-induced} contribution $\Brac{\delta \vecP{\vec u}{i}^{J}/\delta t}$.

The \textit{stochastic} orientations are drawn from the canonical distribution of the Maier-Saupe mean-field approximation $f_{\text{ori}}\left(\vec{u}_i\right) = f_0 \exp \Brac{Us_c \Brac{\vecP{\vec u}{i} \cdot \vecP{\vec n}{c}}^2/\kbt}$, centered around $\vecP{\vec n}{c}$, with normalisation constant $f_0$ and mean field interaction constant $U$~\cite{shendruk2015multi}. 
The local director $\vecP{\vec n}{c}$ is the eigenvector of the cell's tensor order parameter $\tens{Q}_c\Brac{t} = \frac{1}{d-1} \av{d\vecP{\vec u}{i}\Brac{t}\vecP{\vec u}{i}\Brac{t} - \tens{1}}_{c}$, where $\tens{1}$ is the identity tensor in $d$ dimensions.
This eigenvector is associated with the largest eigenvalue $s_c$, which is the scalar order parameter.
The global nematic interaction constant $U$ governs the alignment strength between the MPCD nematogens.

The \textit{flow-induced} reorientations arise from the response of MPCD nematogens to velocity gradients. 
These orientation changes are captured through Jefferey's equation 
\begin{align}
    \frac{\delta {\vecP{\vec u}{i}}^J}{\delta t} &= \alpha \left[ \vecP{\vec u}{i} \cdot \tens{\Omega}_c + \lambda \Brac{\vecP{\vec u}{i} \cdot \tens{E}_c - \vecP{\vec u}{i}\vecP{\vec u}{i}\vecP{\vec u}{i}\colon \tens{E}_c} \right] , 
\end{align} 
for a bare tumbling parameter $\lambda$ and hydrodynamic susceptibility $\alpha$ in a flow with strain rate tensor $\tens{E}_c\Brac{t} = \Brac{\nabla \vecP{\vec v}{c} + \Brac{\nabla \vecP{\vec v}{c}}^T}/2$ and rotation rate tensor  $\tens{\Omega}_c\Brac{t} = \Brac{\nabla \vecP{\vec v}{c} - \Brac{\nabla \vecP{\vec v}{c}}^T}/2$. 
The rotations of nematogens generate hydrodynamic motion, known as backflow, which is accounted for by the change in angular momentum 
$\delta\vec{\mathcal{L}}_{\text{ori}} = - \delta t \Brac{\gamma_R \sum_{j=1}^{N_c(t)} \vecP{\vec u}{i}\Brac{t} \times \vecP{\vec {\dot{u}}}{i}}$, where $\gamma_R$ is the rotational friction coefficient and $\vecP{\vec {\dot{u}}}{i} = \Brac{\delta \vecP{\vec u}{i}^{ST}/\delta t} + \Brac{\delta \vecP{\vec u}{i}^{J}/\delta t}$.

\subsection{Molecular Dynamics}
\label{sctn:md}

The flexible polymer is simulated using a molecular dynamics (MD) technique~\cite{hospital2015molecular}, consisting of $N$ beads with mass $M$, interacting through bond and excluded volume potentials.
Each monomer $j$ obey the equation of motion
\begin{align}
    \label{eq:EquationOfMotion}
    M\vecP{\Ddot{\vec r}}{j} = - \nabla V_j+
    \vecP{\vec \VCO}{j,c},
\end{align}
in which $V_j$ is the total potential of the particle $j$ and $\vecP{\vec \VCO}{j,c}$
represents the momentum exchange between MD beads and fluid particles.
This term accounts for thermal, active, and hydrodynamic drag forces by including MD particle $j$ in the MPCD collision of cell $c$. 
The beads are connected by a finitely extensible nonlinear elastic (FENE) bond potential~\cite{Grest1986,Kremer1990,Slater2009}:
\begin{align}
    \label{eq:FENE}
    V_{\text{FENE}}(r_{jl}) = -\frac{k_\text{FENE}}{2} {r_0}^2 \ln\left(1 - \frac{r_{jl}^2}{{r_0}^2}\right),
\end{align}
where $k_\text{FENE}$ is the bond strength, $r_0$ the maximum bond length and $r_{jl} = \abs{ \vec{r}_j - \vec{r}_l }$ the distance between monomers $j$ and $l$, with $l = j - 1$ for bonded pairs.
Excluded volume effects are modeled using a purely repulsive variant of the Lennard-Jones potential, known as the Weeks-Chandler-Andersen (WCA) potential~\cite{royall2024colloidal,Weeks1971}
\begin{equation}
    \label{eq:WCA}
    V_{\text{WCA}}(r_{jl}) = 4\epsilon 
    \begin{cases}
          \left(\frac{\sigma}{r_{jl}}\right)^{12} - \left(\frac{\sigma}{r_{jl}}\right)^6 + \frac{1}{4}, & r_{jl}<\sigma_{\text{co}}\\
         0, & r_{jl} \geq \sigma_{\text{co}},
    \end{cases}
\end{equation}
where \(\epsilon\) controls the strength of the repulsive potential and \(\sigma\) represents the effective size of a bead.
The potential smoothly approaches zero when the distance between beads $r_{jl}$ exceeds the cutoff distance $\sigma_{\text{co}} = 2^{1/6}\sigma$.
We employ the velocity-Verlet algorithm to numerically integrate~\eq{eq:EquationOfMotion}.

\subsection{Simulation units and parameters}
\label{sctn:unitsParameters}

The MPCD cell size $a$ sets the unit of length, thermal energy $\kbt$ defines the energy unit and the mass unit is fluid particle mass $m$ (with all fluid particles having the same mass $m_i = m$).  
These units determine other simulation units, including units of time $t_0 = a\sqrt{m/\kbt}$.
The time step is set to $\Delta t = 0.1 t_0$.
The average number of AN-MPCD particles per cell is $\av{N_c} = 20$, where $\av{\cdot}$ denotes the average over the entire system. The nematic interaction constant $U = 10\kbt$ is deep within the nematic phase~\cite{shendruk2015multi}.
The fluid particles are initialized with random positions, Maxwell-Boltzmann distributed velocities, and uniform orientations along the $\hat{x}$-axis.
The rotational friction coefficient is $\gamma_\text{R} = 0.01\kbt t_0$, hydrodynamic susceptibility  $\alpha = 0.5$ and tumbling parameter is $\lambda = 2$ for a flow-aligning liquid crystal~\cite{shendruk2015multi}.
We explore a range of activity, $\zeta \in \{3, \ldots, 8\}\times 10^{-2}$, with $\sigma_p = 0.5 $ and $\sigma_w = 0.1$ in activity modulation function. 
These values give optimal density behavior without lowering effective activity~\cite{kozhukhov2024mitigating}.

The MD bead size is $\sigma  = 1a$ and the repulsive strength is $\epsilon = 1\kbt$.
The maximum bond length is $r_0 =1.5a$, and the bond strength is $k_\text{FENE}=120\kbt/a^2$.
Monomer mass is $M =10m$.
The degree of polymerization is $N={5,10,15,20,25,50,100,200}$, corresponding to a contour length $ = (N-1) b$, where $b$ is the average bond length.
The MD time step is ${\Delta t}_{\text{MD}} = 0.002t_0$, requiring $50$ MD iterations per AN-MPCD iteration.

\begin{figure}[tb]
    \centering
    \begin{subfigure}
        \centering
        \includegraphics[width=0.89\linewidth]{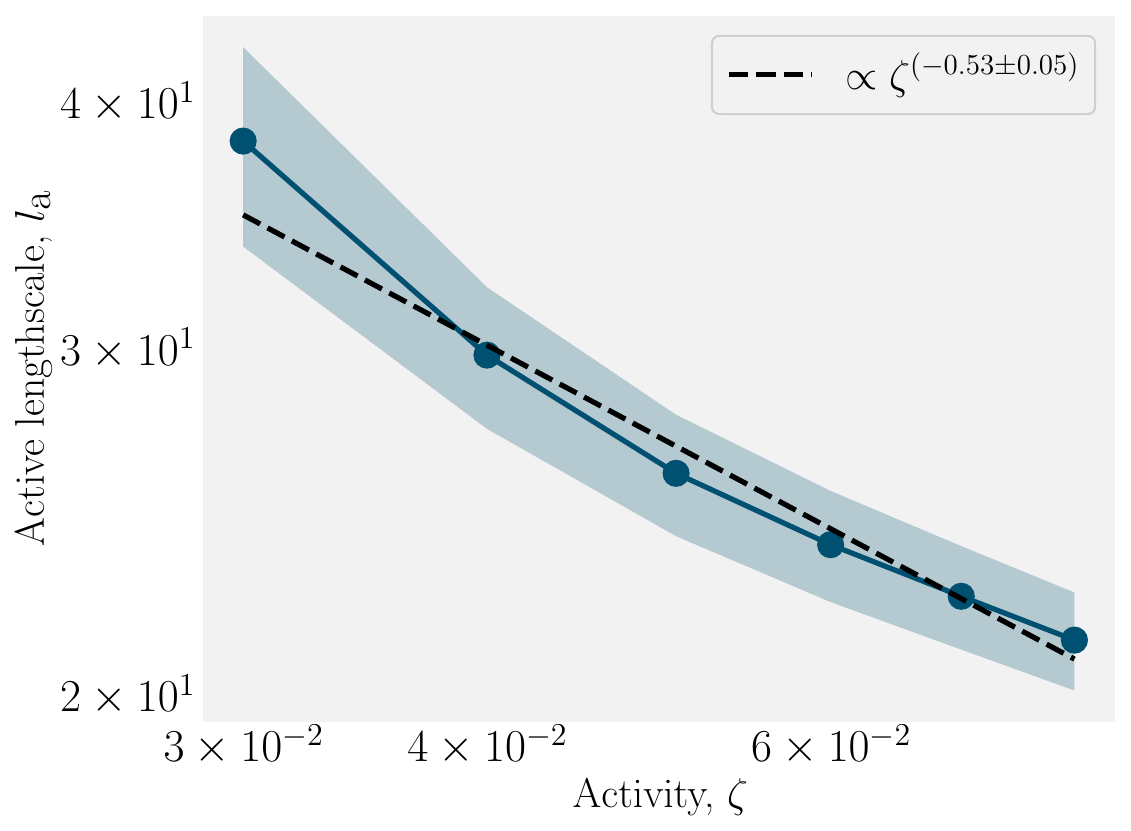}
    \end{subfigure}
    \begin{subfigure}
        \centering
        \includegraphics[width=0.89\linewidth]{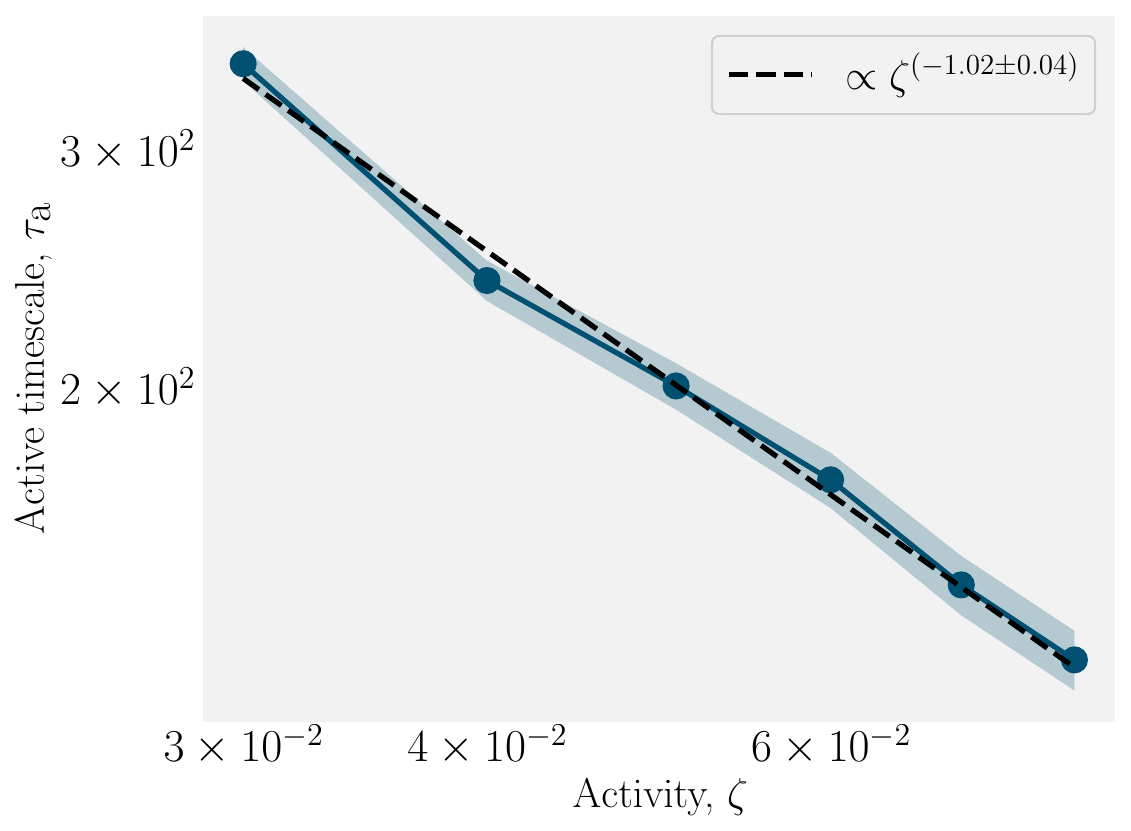}
    \end{subfigure}
    \caption{\textbf{Characteristic scales of active turbulence in AN-MPCD.} 
    (a) Active length scales as measured from the square root of system size divided by average number of all defects, with the shaded area representing the standard deviation of the number of defects.
    (b) Active time scale as measured from an exponential fit to the temporal velocity-velocity autocorrelation function, with shaded area showing the error on the fit. 
    The dashed lines indicate power law fits.}
    \label{fig:ActiveScales}
\end{figure}

Polymers are embedded in a periodic box of size $l_\text{sys}\times l_\text{sys} = 220 a \times 220 a$ with periodic boundary conditions, unless stated otherwise.
The polymer is initialized in a fully extended conformation, aligned with the global nematic direction, centered at the simulation box's center, and allowed to relax.
Data is recorded once the system reaches its steady state, which is identified via an iterative procedure.
In each iteration, the ensemble average is compared to the overall average calculated from all repetitions to find the first instance where it falls below this overall average.
This time is then used as the starting point for updating the overall average.
The process is repeated until successive updates of this reference time do not alter and stabilize, indicating that the system has achieved a steady state.
Forty repeats for each set of parameters are simulated, each lasting $2.2-2.5\times 10^4 t_0$.

\subsection{Mean-Squared Displacement}
\label{sctn:msd}

To understand the MSD of the polymers in an active nematic, we can treat the center of mass $\rcmV=N^{-1}\sum_i^N\posV_i$ as a persistent random walker with a passive translational diffusion coefficient $\diffTrans = R_g^2/\timeRelax$, constant instantaneous speed $\actSpeed$ and a persistence time $\timePers$ for the direction of motion $\hat{e}$. 
The equations of motion are
\begin{subequations}
\begin{align}
    \frac{d \rcmV}{d \time} &= \actSpeed \ori + \sqrt{2 \diffTrans} \noiseV_\text{\pos} \\
    \frac{d \ori}{d \time} &= \sqrt{2 \timePers^{-1} } \noiseV_{\ori} \times \ori, 
\end{align}
\end{subequations}
where noises are white with zero mean $\av{\noiseV_i(\time)} = 0$ and unit variance $\av{\noiseV_i(\time) \ \noiseV_i(\time')} = \unitT \delta(\time-\time')$ for both the translational and orientational dynamics, $i \in \left\{ \text{\pos}, \ori \right\}$. 

In the passive limit, the persistence time and the speed go to zero, such that the dynamics are entirely determined by the translational diffusion coefficient.
When activity is nonzero, the motion is modeled by Langevin equations for the center of mass position and orientation, and the MSD is~\cite{Zottl2016}
\begin{align}
    \av{\delta \rcmV^2} &= 2d \diffTrans\time + 2\actSpeed^2\timePers\time - 2\actSpeed^2\timePers^2 \left[ 1 - e^{-\time/\timePers} \right].
    \label{eq:langevinModel}
\end{align}
The MSD of center-of-mass of polymers agrees well with this model (\fig{fig:MSD}c; dashed line).
By fitting our simulation data for MSD to this model, we extract the short time speed of the center of mass of the polymers $\actSpeed$ and the persistence time $\timePers$ (\fig{fig:MSD}d,e).

\begin{figure}[tb]
    \centering
    \includegraphics[width=0.89\linewidth]{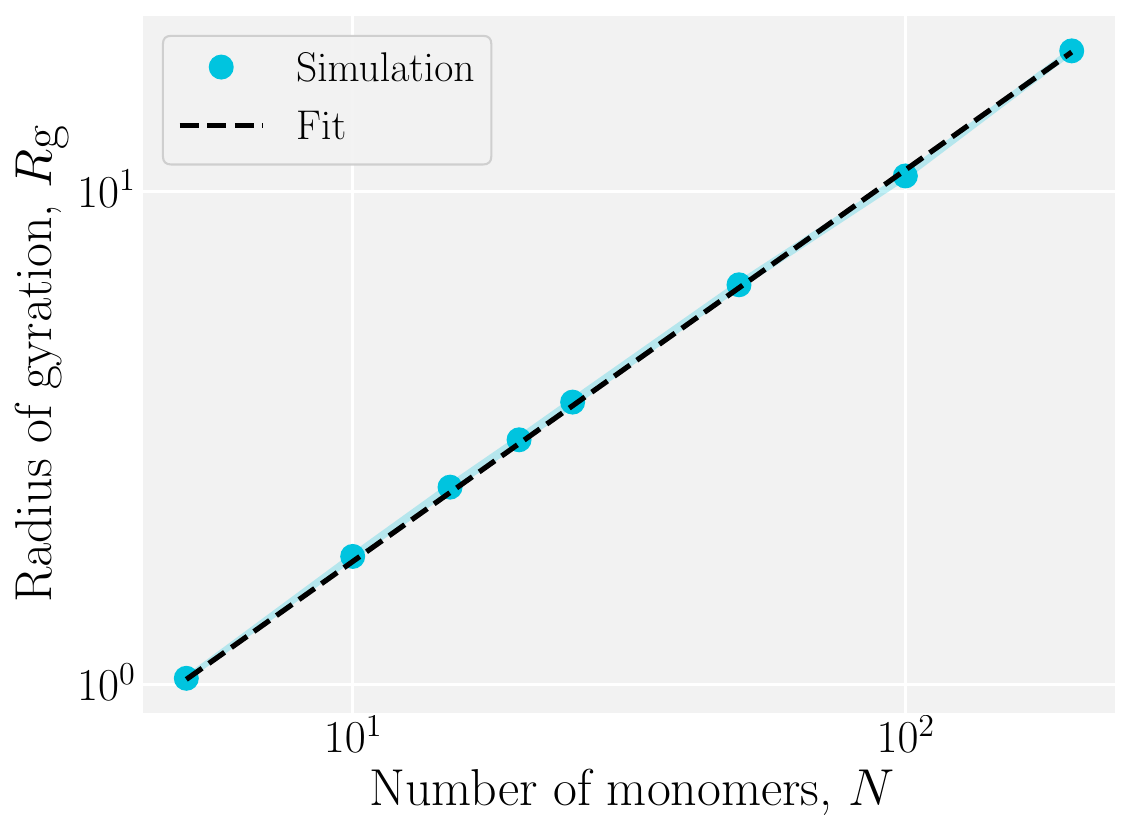}
    \caption{
    \textbf{Gyration radius scaling.}
    Under good solvent conditions and in the presence of hydrodynamic and excluded volume interactions, the gyration radius scales as $R_g \sim N^{3/4}$ in 2D. 
    Shaded area shows the standard deviation. 
    The dashed line indicates the fit to $R_g \sim N^\nu$, where $\nu = 0.77 \pm 0.01$.
    }
    \label{fig:radiusGyrations}
\end{figure}

Taking the long-time limit of \eq{eq:langevinModel} ($t \gg \timePers$), results in $ \av{\delta \rcmV^2} = 2d \diffTrans\time + 2\actSpeed^2\timePers\time$, with an effective diffusion coefficient given by~\eq{eq:effDiff}.    
The effective diffusion coefficients shown in \fig{fig:MSD}f are calculated by substituting the fitted values for $\actSpeed$ and $\timePers$ from~\eq{eq:langevinModel} into~\eq{eq:effDiff}.

\subsection{Characterizing Active Turbulence in Active Nematic Multi-Particle Collision Dynamics}
\label{appendix:turb}

The scaling arguments presented in the main text rely on characteristic scaling in active turbulence. 
In particular, active turbulence has a characteristic length scale that arises due to the interplay of elasticity and activity, as $l_a \sim \sqrt{K/\act} \sim \act^{-1/2}$. 
The expected scaling is reproduced in 2D AN-MPCD simulations, which reproduce a power law dependence of $l_a \sim \act^{-0.53\pm0.05}$ (\fig{fig:ActiveScales}a). 
Similarly, the competition between viscosity and activity defines an active time scale that is expected to scale as $\tau_a \sim \eta/\act \sim \act^{-1}$. 
The AN-MPCD simulations agree with this expectation, reproducing a scaling of $\tau_a \sim \act^{-1.02\pm0.04}$ (\fig{fig:ActiveScales}b). 

\begin{figure}[tb]
    \centering
    \includegraphics[width=0.89\linewidth]{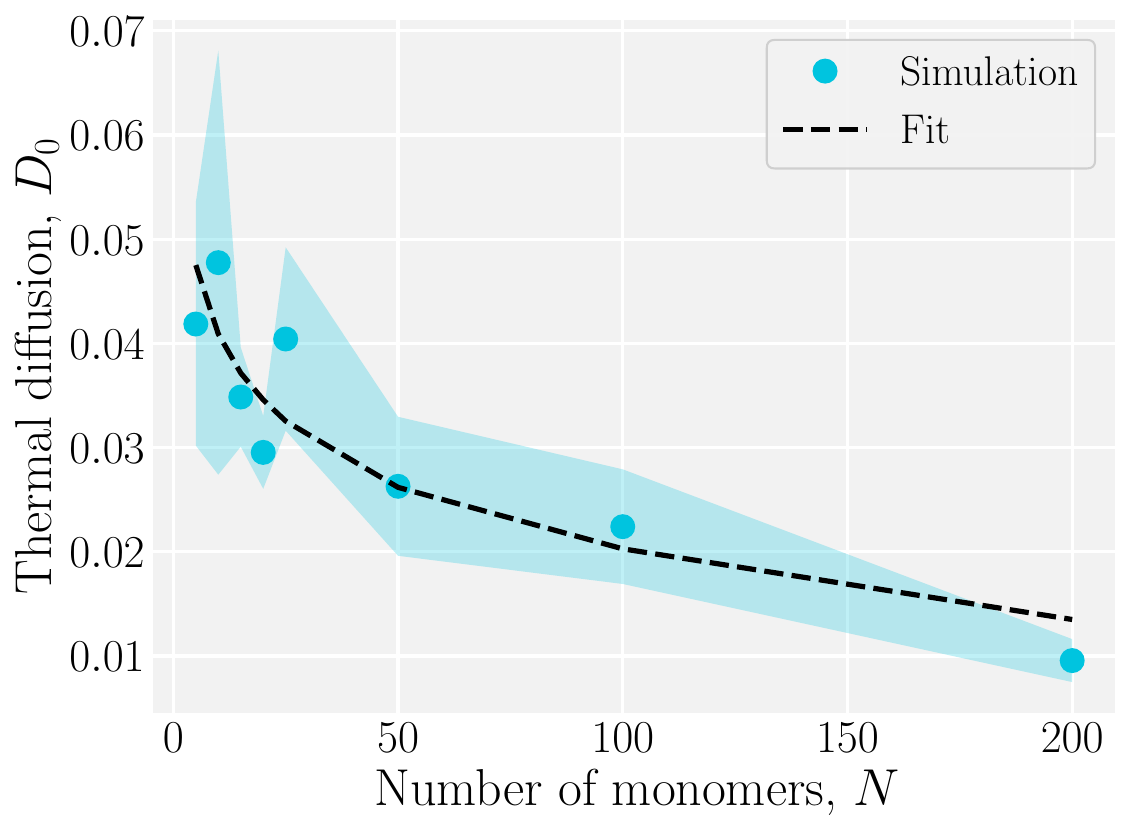}
    \caption{
    \textbf{Thermal diffusion coefficient scaling.} 
    The diffusion coefficient scales as $\diffTrans \sim \ln\Brac{L/R_G}$ in 2D.
    The dashed line indicates the fit to \eq{eq:diff2D}.
    Dots show the simulation results, and shaded area shows the standard deviation. 
    }
    \label{fig:thermDiff}
\end{figure}

\section{Passive polymer properties}
\label{sctn:passive}

In the dilute limit, the gyration radius $R_g$ and the diffusion coefficient of the polymer's center of mass $D_0$, as functions of the number of monomers $N$, are expected to follow the scaling relations
\begin{align}
    R_g &\sim N^\nu,
\end{align}
with scaling exponent $\nu$. 
Without hydrodynamic interactions and excluded volume effect, the simple Rouse model gives $\nu = 1/2$.
For dilute systems in 3D, excluded volume effects result in $\nu \approx 3/5$.
In the presence of excluded volume effect in 2D, $\nu = 3/4$.
The scaling of the gyration radius for passive polymers in our simulations is $\nu = 0.77 \pm 0.01$, in close agreement with the expectation value of $3/4$ (\fig{fig:radiusGyrations}).

While intuition from 3D systems may lead us to expect the thermal diffusion to scale as a power law with the polymer length, extensive computational finite-size scaling studies have demonstrated that, in 2D, the diffusion coefficient follows the form $\diffTrans \sim \ln{\left( l_\text{sys} / R_g \right)} \sim \mathcal{D}^{-1}_N$, where $l_\text{sys}$ is the linear size of the system~\cite{falck2003dynamics,punkkinen2005dynamics}. 
Our simulations results exhibit good concordance with the logarithmic form of 
the passive diffusion coefficient, with the expected logarithmic behavior falling within the standard deviation of these measurements except for $N=200$ (\fig{fig:thermDiff}).

\section{Preliminary \peclet Number Scaling}
\label{sctn:preliminaryPe}

\begin{figure}[tb]
    \centering
    \includegraphics[width=0.89\linewidth]{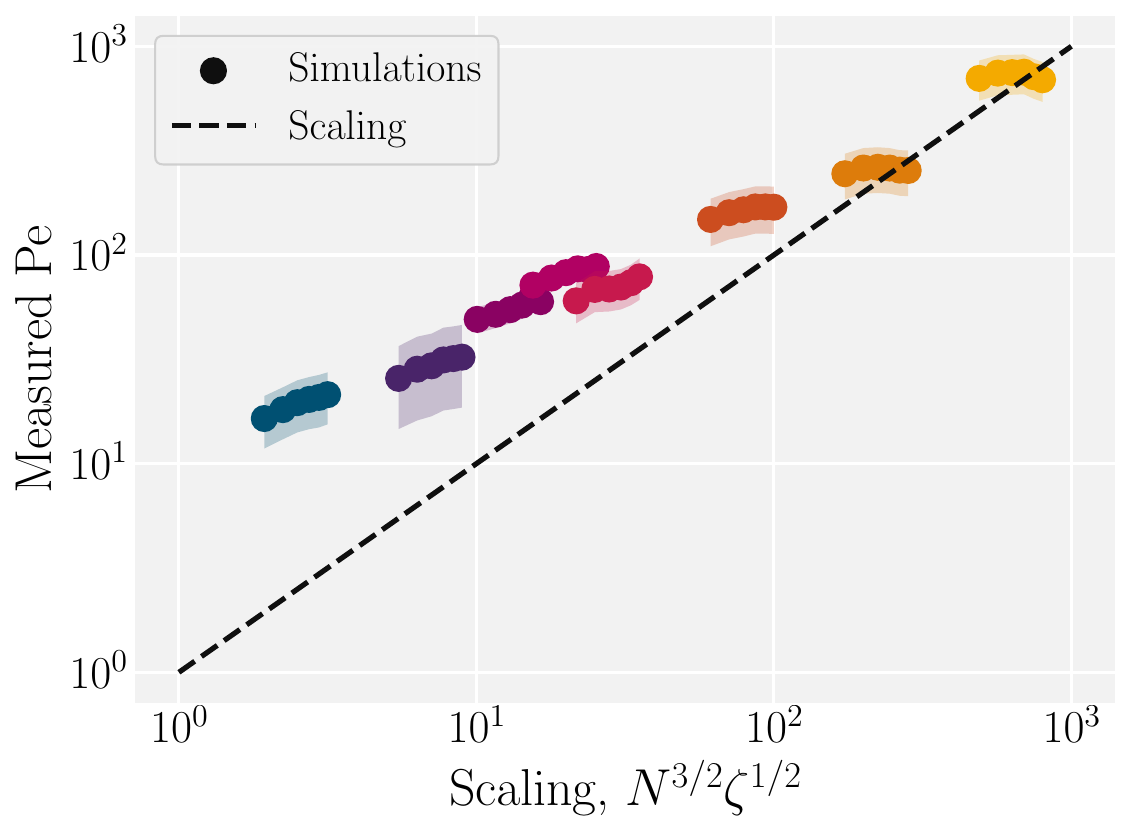}
    \caption{
    \textbf{Scaling of the \peclet number.} 
    The \peclet number $\pe=v R_g/\diffTrans$ is measured directly from simulations with velocity $v$ from the MSDs (\fig{fig:MSD}) and $R_g$ and $\diffTrans$ from passive simulations (\sctn{sctn:passive}). 
    The naive scaling of $\pe\sim N^{3/2}\act^{1/2}$ with length $N$ and activity $\act$ is discussed in the text and seen to be erroneous.
    }
    \label{fig:3DPeclet}
\end{figure}

In the following, we present a naive scaling argument for $\pe$, which we will show fails for polymers in 2D finite active turbulence. 
The passive diffusion coefficient might be erroneously estimated by the Stokes-Einstein relation as $D_0 \sim 1/R_g \sim N^{-3/4}$, insterad of using the logarithmic form of \eq{eq:diff2D}.
Additionally, one might reasonably hypothesize that the polymer's speed $v$ is only proportional to the characteristic active flow speed $v_a$, which itself goes as $v_a \sim l_a / \tau_a \sim \act^{1/2}$
by~\eq{eq:actLengthTime}~\cite{hemingway2016correlation}, which neglects the crossover behavior of \eq{eq:crossover}. 
By substituting the scaling expectations for $R_g\sim N^{3/4}$, $D_0 \sim N^{-3/4}$ and $v_a\sim\act^{1/2}$ into \eq{eq:pe}, one would predict $\pe \sim N^{3/2}\zeta^{1/2}$.
However, the resulting scaling fails to describe the measured \peclet number (\fig{fig:3DPeclet}).

\begin{figure}[b]
    \centering
    \includegraphics[width=0.89\linewidth]{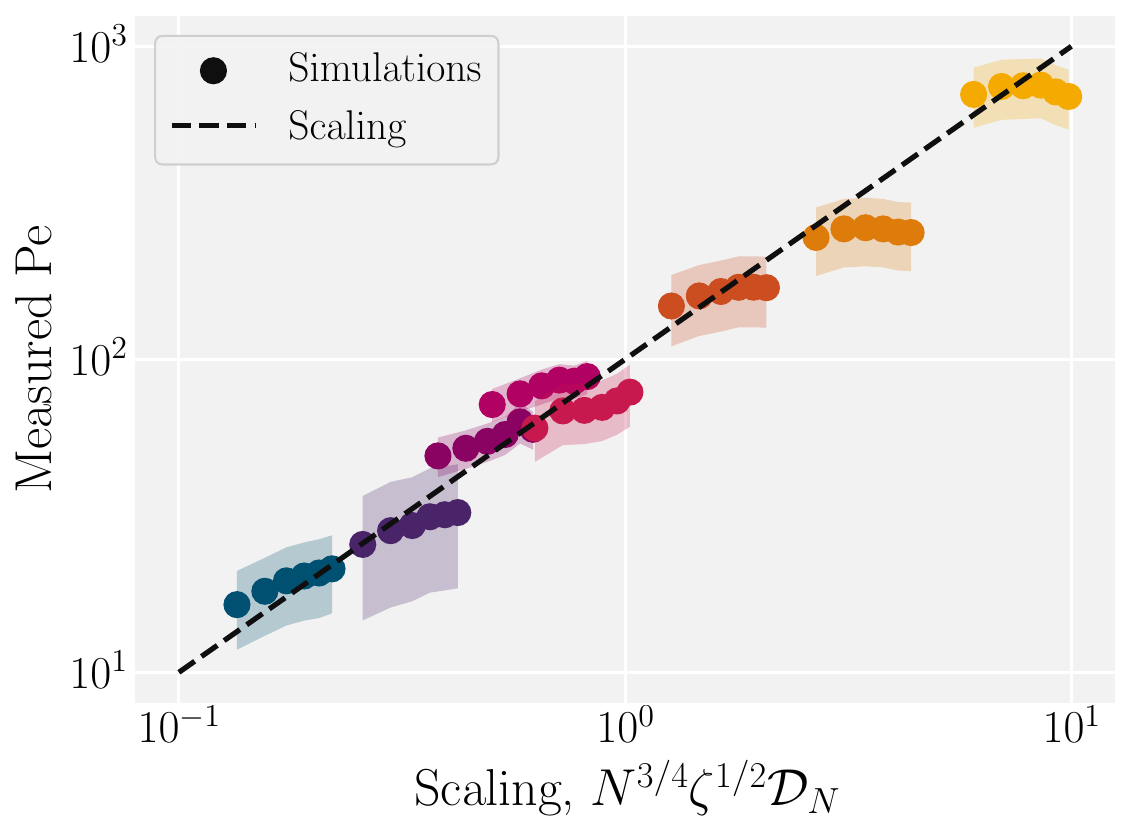}
    \caption{
    \textbf{Scaling of the \peclet number.} 
    Same as \fig{fig:3DPeclet} but the \peclet number $\pe=v R_g \mathcal{D}_N$ uses the 2D thermal diffusivity $\mathcal{D}_N^{-1}$ from \eq{eq:diff2D}. 
    }
    \label{fig:logDiff}
\end{figure}

Since the gyration radius behaves as expected (\sctn{sctn:passive}), we rule out its scaling as the source of the deviation. 
However, the expectation that $\diffTrans\sim R_g^{-1}$ is incorrect in 2D (\sctn{sctn:passive}).
Plotting $\pe$ against $N^{3/4}\act^{1/2}\mathcal{D}_N$ demonstrates that this resolves the primary issue with the scaling (~\fig{fig:logDiff}). 

While the scaling holds at short contour lengths, it begins to saturate with activity at long lengths, 
demanding the crossover scaling for the polymer speed as discussed in the main text. 

\end{document}


\section{Active Nematic Multi-Particle Collision Dynamics}
\label{appendix:methods}

\begin{figure*}
    \centering
    \includegraphics[width=0.6\linewidth]{images/Lact2.pdf}
    \caption{Active length scale. The interplay of elasticity and activity defines an active length scale $l_a$, theoretically scaling as $\propto \zeta^{-\frac{1}{2}}$. 
    Dots represent simulation values, with shaded area showing standard deviations.
    The dashed line indicates the fit.}
    \label{fig:ActiveLength}
\end{figure*}
\begin{figure*}
    \centering
    \includegraphics[width=0.6\linewidth]{images/Tact-flow5.pdf}
    \caption{Active time scale. The competition between viscosity and activity defines a active time scale $\tau_a$, theoretically scaling as $\propto \zeta^{-1}$. 
    Dots represent simulation values, with shaded area showing standard deviations.
    The dashed line indicates the fit.}
    \label{fig:ActiveTime}
\end{figure*}

\bibliographystyle{unsrt}
\bibliography{references}